\documentclass[a4paper,twocolumn,showpacs,superscriptaddress,prl]{revtex4}  
\usepackage{epsfig}

\begin{document}

\newcommand{\vect}[1]{{\bf #1}}
\newcommand{\comments}[1]{\hfill {\tt {eq:~#1}}} 

\title{Non-linear quantum critical transport and the Schwinger Mechanism}

\author{A. G. Green}
\address{School of Physics and Astronomy, University of St Andrews,
North Haugh, St Andrews KY16\ 9SS, UK}

\author{S. L. Sondhi}
\address{Department of Physics, Princeton University, Princeton, NJ 08544, USA}

\date{\today}

\begin{abstract}
Scaling arguments imply that quantum critical points exhibit universal
non-linear responses to external probes. We investigate the origins of
such non-linearities in transport, which is especially problematic since
the system is necessarily driven far from equilibrium. We argue that
for a wide class of systems the new ingredient that enters is the
Schwinger mechanism---the production of carriers from the vacuum by
the applied field--- which is then balanced against a scattering rate
which is itself set by the field. We show by explicit computation how
this works for the case of the symmetric superfluid-Mott insulator transition
of bosons.
\end{abstract}

\maketitle

The notion of quantum criticality provides one the few general  
approaches to the study of strongly correlated quantum many-body
systems\cite{Sachdev1999}. The scale invariance that characterizes the zero
temperature critical point leads to characteristic universal power law
dependences for various quantities in its
proximity; these
dependences can be computed within a continuum field theory

While the power law dependences can generally be related to an
underlying set of (possibly unknown) exponents by scaling arguments,
establishing the actual mechanism that gives rise to them--- in the
sense of prescribing a controlled computation--- is not always a
trivial task. For example, much attention has focussed on
understanding the 
real-time dynamics at non-zero temperatures\cite{Damle1997} where 
the textbook procedure of 
analytic continuation from Matsubara expressions typically
yields incorrect answers and no insight.

Here, we address another such question, that of transport
at finite fields. Specifically, consider a quantum critical point
between an insulator and a metal or superconductor/superfluid
characterized by
a correlation length that diverges as $\xi \sim \delta^{-\nu}$ and an
energy scale that vanishes as $\Delta \sim \delta^{\nu z}$, where
$\delta$ is a measure of the distance to the transition. Dimensional
analysis implies that the zero-temperature conductivity obeys the scaling form
\begin{equation}
\sigma(\delta, E) 
= 
E^{d - 2 \over z + 1} 
\Sigma \left( 
\frac{\delta }{E^{\nu(z+1)}} 
\right)
\label{Scaling}
\end{equation}
in $d$ spatial dimensions.
Thence, generically, the system exhibits a non-linear current-voltage
characteristic at criticality. The question of interest to us is how
the system might produce such a response. Evidently, linear response theory
or naive perturbation theory to higher orders is no help. Physically,
the system must set up a steady state whose properties depend in singular
fashion on $E$ with no expectation that it resembles the steady state
obtained in thermal equilibrium.

We show that the properties of such weakly interacting fixed-points
may be understood as follows:
The application of an electric field leads to a biased growth of current 
carrying fluctuations by an 
analogue
of the Schwinger mechanism\cite{Kluger1991}, by which an electric
field produces electron/hole 
pairs from the vacuum. This process is non-perturbative in the electric
field. The fluctuations produced in this way scatter from one another due 
to the interactions at the fixed point, thus producing current relaxation 
which is again non-perturbative in the electric field at the fixed
point.
Together, these effects establish a steady-state distribution that carries
a current.
We implement this idea for the simplest quantum phase-transition with
singular transport--- the symmetric superfluid to Mott insulator transition of
bosons in a periodic potential--- and recover the dependence
(\ref{Scaling}) with $z=1$. 
In particular, in $d=2$ we find a linear conductivity 
$\sigma= (\pi/4)e^{*2}/h$, different from those calculated previously \cite{Cha1991,Damle1997}.
The difference between these results arises due to the different regimes of frequency, temperature and electric field ($\omega$,$T$,$E$) to which they apply and the different physical processes important in each case. The linear response calculated at zero temperature\cite{Cha1991} amounts to the limit ($\omega \rightarrow 0$, $T=0$, $E=0$), whilst the finite temperature, linear response\cite{Damle1997} corresponds to ($\omega=0$, $T \ne0$, $E=0$). Our result follows from taking the frequency and temperature to zero at finite field; ($\omega=0$, $T=0$, $E\ne 0$). The non-commutativity of the limits $\omega \rightarrow 0$ and $T \rightarrow 0$ was first recognized in\cite{Damle1997}. The result presented here suggests that similar care should be taken with the electric field. Next, we outline our 
computation which builds upon the important work by Damle and
Sachdev\cite{Damle1997} on the  
finite-temperature transport of the same model\cite{Dalidovich2004}.

\paragraph*{Field Theory}
The critical region of the superfluid to Mott insulator transition with particle hole symmetry is described by
a charged scalar field with a quartic interaction\cite{Sachdev1999,Damle1997}: 
\begin{equation}
{\cal H}
=
\int d^d x 
\left[
\Pi^{\dagger} \Pi + \nabla \phi^{\dagger} \nabla \phi +m^2 \phi^{\dagger} \phi
+
\lambda (\phi^{\dagger} \phi)^2
\right],
\label{Hamiltonian}
\end{equation}
where $\phi$ is the complex scalar field and $\Pi$ is its conjugate
momentum. These satisfy the usual commutation relations 
$
\left[ \phi({\bf x},t), \Pi({\bf y},t) \right] 
=
i \delta({\bf x}-{\bf y})
$.
It is convenient to choose the bare interaction $\lambda$ to have its
fixed point value $u^* \Lambda^{3-d}$ with momentum cutoff $\Lambda$
\cite{Sachdev1999} although
we will not need the precise, regularization-dependent, value.
At the
zero-temperature critical point, the renormalized mass is zero which
corresponds to a particular choice $m^*$ of the bare mass.
The effects of applying an electric field, ${\bf E}$, are included by
minimally coupling to a vector potential; $\nabla \phi \rightarrow D
\phi=\left(\nabla +i {\bf A} \right)\phi$. We choose the gauge ${\bf
  A}={\bf E}t$ for initial convenience, later we switch to a scalar
potential. 

\paragraph*{ Mode Expansion}
The normal modes of this Hamiltonian (in the absence of interaction)
are charge density fluctuations. These occur with positive and negative
charges, corresponding to decrease or increase in charge density from
the average. The first step in our analysis is to re-express 
(\ref{Hamiltonian}) in terms of creation and annihilation operators 
for these normal modes. The transformation is standard for the
Klein-Gordon theory. Using $a^{\dagger}$ and $a$ to represent the
creation and annihilation of positively charged density fluctuations
and $b^{\dagger}$ and $b$ for negatively charged fluctuations, the
non-interacting part of the Hamiltonian reduces to
\begin{eqnarray}
{\cal H}_0
&=&
\int \frac{d^2k}{(2 \pi)^2}
\left( a^{\dagger}({\bf k},t), b(-{\bf k},t) \right)
\nonumber\\
& &
\;\;\;\;\;\;\;\;\;\;\;
\times
\left(
\begin{array}{cc}
\epsilon_{\bf k} + B_{\bf k} & B_{\bf k}\\
B_{\bf k} & \epsilon_{\bf k} + B_{\bf k}\\
\end{array}
\right)
\left(
\begin{array}{c}
a({\bf k},t)\\
b^{\dagger}(-{\bf k},t)\\
\end{array}
\right),
\label{Hamiltonian2}
\end{eqnarray}
where $\epsilon_{\bf k}=\sqrt{ {\bf k}^2+m^2}$ is the mode energy and 
$B_{\bf k}=\left( ({\bf E}t)^2+2 {\bf k}.{\bf E}
  t\right)/2\epsilon_{\bf k}$. The Hamiltonian has an
explicit time dependence arising from our choice of gauge used in
coupling to the electric field. This time dependence is responsible for the
production of fluctuations by the electric field.

\paragraph*{ Schwinger Mechanism} 
Let us ignore the interaction between the normal modes of the
system. This will allow us to describe the pair production
process, after which we will put back the
interactions. Our first step is to diagonalize the Hamiltonian
(\ref{Hamiltonian2}) using a 
Bogoliubov transformation. Because of the time dependence of the
Hamiltonian, the transformation used to carry out this diagonalization 
must itself be time dependent. This has important consequences for the
equations of motion. In the instantaneously diagonal basis, or {\it
  adiabatic particle basis}, the Heisenberg equations of motion for
operators pick up extra terms from the time dependence of the
Bogoliubov transformation matrix.  
We are concerned with the equations of motion for the 
regular and anomalous distribution functions;
$
f({\bf k},t)=\langle a^{\dagger}({\bf k},t)a({\bf k},t) \rangle
=\langle b^{\dagger}(-{\bf k},t)b(-{\bf k},t) \rangle
$
and
$g({\bf k},t)=\langle b(-{\bf k},t)a({\bf k},t) \rangle$, where the
angular brackets indicate averages over the state of system\cite{Distributions}. After
transforming to the adiabatic particle basis, the
equations of motion for these reduce to
\begin{eqnarray}
\frac{df({\bf k},t)}{dt}
&=&
\frac{\dot \epsilon_{\bf k}(t)}{\epsilon_{\bf k}(t)} {\cal R}e g({\bf k},t)
\nonumber\\
\frac{dg({\bf k},t)}{dt}
&=&
\frac{\dot \epsilon_{\bf k}(t)}{2\epsilon_{\bf k}(t)}(2f({\bf k},t)+1 )
-2i \epsilon_{\bf k}(t) g({\bf k},t)
\label{f_g_motion}
\end{eqnarray}
where $\epsilon_{ \bf k}(t)=\sqrt{(\epsilon_{\bf k} + B_{\bf k})^2-B_{\bf k}^2}
=\epsilon_{{\bf k}+{\bf E}t}$ is the mode energy in the adiabatic
particle basis.  
These equations contain all of the ingredients necessary to describe
the Schwinger mechanism\cite{Kluger1991}. The terms proportional to
$\dot \epsilon_{\bf k}(t)$ 
result from the time dependence of the Hamiltonian and are responsible
for pair creation. The second equation can be solved explicitly
for $g({\bf k},t)$ and substituted back into the first. The resulting
equation contains a source term for the production of
fluctuations, which includes oscillatory behavior coming from the quantum coherence of the pair production. If we ignore these transients, the source term may be replaced by a delta-function with the appropriate
weight. The equation of motion for $f({\bf k},t)$ then reduces to 
\begin{equation}
\frac{df({\bf k},t)}{dt}
=
\delta(t+{\bf k}.{\bf E}/{\bf E}^2)
e^{-\pi (k \perp^2+m^2)/E}.
\label{Source2}
\end{equation}
The pair production described in this way may be understood by analogy
with Landau-Zener tunneling\cite{Zener1932} as follows: under the action of
the electric field, the momentum of a charged excitation increases
linearly in time as ${\bf k}+{\bf E}t$. The energy of the charged
excitation changes in time accordingly; at large momenta it is
proportional to $|{\bf k}+{\bf E}t|$, and at a time 
$t=-{\bf k}.{\bf E}/E^2$ it passes though a minimum equal to $m$. This
is very similar to the energy dependence of modes in the Landau-Zener
model. Except for the Bose enhancement factor $2f({\bf k},t)+1$ the
pair production equations (4) have precisely the same form as in Landau-Zener
tunneling.

\paragraph*{Scattering and Steady State}
The steady creation of pairs from the vacuum will,  in the absence of
scattering, 
lead to a secular divergence of the current. Thus, consideration of
the scattering is essential for understanding
the steady state transport.  Below $d=3$, the critical behavior is controlled
by the interacting, Wilson-Fisher fixed point. If we access the
properties of this fixed point in a weak coupling expansion, such
as the $1/N$ expansion, which we use in this paper, then the leading
order description of the critical, steady state can be obtained by
considering the scattering of the particles produced via the
Schwinger mechanism. 
Heuristically, the pair production in (\ref{Source2}) leads to a growth in the
current with a rate proportional to $E^{d + 1 \over 2}$, while
scattering with a coupling of order $1/N$ to $N$ channels
is expected (on dimensional
grounds) to lead to a current relaxation rate of order $(1/N)
\sqrt{E}$; together these will reproduce (\ref{Scaling}) with 
$z=1$. In the remainder of the paper we shall see how this
insight can be turned into a computation within the quantum
Boltzmann equation framework. 

Generally, $1/N$-expansions extend the number of modes of the model 
from its initial value to some large number $N$, allowing all of these modes
to interact with one another. If $N$ is taken to be very large, the
interaction may be expanded perturbatively in $1/N$.
Crucially, in the present case, we couple the electric field to just two modes of our extended
theory ({\it i.e.} one of the $N/2$ copies of the model). The field induces fluctuations in these two modes   {\it via} the Schwinger mechanism. These fluctuations scatter into the
remaining $N-2$ modes, thus allowing the modes that are coupled to the
electric field to reach a steady state. We discuss moving beyond
leading order later in the paper.  The details of how this $1/N$ expansion
is set up are given in \cite{Sachdev1998}.

The scattering integrals in our Boltzmann equations may be determined
using several methods. In the case where only regular distribution
functions are required, it is possible to use Fermi's Golden
rule. Since we also have to consider anomalous distributions, this
simplest approach does not work. It is, however, possible to determine the
scattering integrals by using Heisenberg's equations of motion with
the $1/N$ Hamiltonian and time-dependent perturbation theory; this is
after all how Fermi's Golden Rule is derived in the first place.

\paragraph*{1-Loop Scattering Integrals}
The 1-loop correction is particularly simple to calculate. After 
calculating the Heisenberg equations of motion including the
interaction Hamiltonian and then using
mean field theory, one finds that the only modification is to replace
$B_{\bf k}$ with $B_{\bf k}+\Sigma_{\bf k}$, where the self-energy 
$\Sigma_{\bf k}=m(E)^2/(2 \epsilon_{\bf k})$. An expression for the 1-loop
renormalized mass $m(E)$ is given to lowest order in a $1/N$-expansion in Eq. (\ref{SelfConsistency}) below\cite{gapON}.
By making a new Bogoliubov
transformation to the {\it 1-loop renormalized adiabatic particle
  basis} one may reduce the Heisenberg equations for the distribution
functions to the same form as (\ref{f_g_motion}), 
replacing the adiabatic mode energy
$\epsilon_{{\bf k}}(t)$ everywhere with the renormalized mode energy,
$\epsilon_{{\bf k},m}(t)=\sqrt{({\bf k}+{\bf E}t)^2+m(E)^2}$.  Notice
that this modification points to a
feedback between the production of fluctuations and the scattering
between them. Since the rate of production depends exponentially upon
the energy gap, it is suppressed by scattering---although its 
functional dependence on $E$ is not changed. Of course, this scattering 
is not sufficient to establish a steady state and indeed, can be
ignored at leading order in $1/N$ altogether as will be clear below.

\paragraph*{2-Loop Scattering Integrals} 
The next order scattering integral is deduced from the Heisenberg
equations of motion for the various mode operators followed by time
dependent perturbation theory to lowest order in $1/N$. To lowest
order in $1/N$ and at zero-temperature, the dominant scattering is due to
processes where the modes coupled to the E-field 
scatter off one another and into the $N-2$ uncoupled modes of the extended
theory. At this juncture, it is useful to make the change of variables
$
t,{\bf q} \rightarrow t,{\bf k}
$,
where
$
{\bf k}={\bf q}+{\bf E}t
$
is the instantaneous momentum of the quasi-particle. Rather than label
the normal modes by their initial momentum and keep the label fixed
in time, we choose to label them with their instantaneous
momentum. This is equivalent to changing gauge to a scalar potential. With this change of 
variables all explicit time dependence is removed, except for in the
distribution functions. In particular, $\epsilon_{\bf
  q}(t)=\epsilon_{\bf k}=\sqrt{{\bf k}^2+m^2}$. The resulting Boltzmann
equations are  
\begin{eqnarray}
\left[ 
\partial_t+{\bf E}.\partial_{\bf k}
\right]
f({\bf k},t)
&=&
+ \frac{{\bf E}. \partial_{{\bf k}}\epsilon_{\bf k}}{\epsilon_{\bf k}} 
{\cal R}e g({\bf k},t)
\nonumber\\
& &
+\gamma_{\bf k}(t) \left[ {\cal R}e g({\bf k},t) - f({\bf k},t) \right]
\nonumber\\
\left[ 
\partial_t+{\bf E}.\partial_{\bf k}
\right]
g({\bf k},t)
&=&
-2 i \epsilon_{\bf k} g({\bf k},t)
+ \frac{{\bf E}. \partial_{{\bf k}}\epsilon_{\bf k}}{\epsilon_{\bf k}} 
\left( 2f({\bf k},t)+1 \right)
\nonumber\\
& &
-\gamma_{\bf k}(t) f({\bf k},t)
\label{Boltzman}
\end{eqnarray}
with the supplementary definitions
\begin{eqnarray}
\gamma_{\bf k}(t)
&=&
\frac{8}{N} 
\frac{  \sqrt{|{\bf k}| + k_{\parallel} }  }{\epsilon_{\bf k}}
\int \frac{d {\bf k}'}{(2 \pi)^2}
\frac{ \sqrt{|{\bf k}'| + k'_{\parallel} } }{\epsilon_{{\bf k}'}}
f({\bf k}',t)
\nonumber\\
m(E,t)
&=&
\frac{16 \pi}{N} \int \frac{d{\bf k} }{(2 \pi)^2} 
\frac{f({\bf k},t)}{\epsilon_{\bf k}}
+ \Delta
\label{SelfConsistency}
\end{eqnarray}
The damping factor $\gamma({\bf k},t)$ is derived allowing for the
zero-temperature, critical propagation of the $N-2$ uncoupled modes.

\paragraph*{Scaling and Current}
The above equations clearly permit a universal, steady state 
solution whose properties are governed only by the electric field and the
fixed point value of the coupling. Writing the scaling forms
$f/g({\bf k},E) = f/g({\bf k}/\sqrt{E}) \equiv f/g(\tilde k)$,
$m = \tilde m \sqrt{E}$ and $\gamma = \tilde \gamma \sqrt{E})$ 
we see that the Eqs.~(\ref{Boltzman}) reduce to $E$ independent
equations for $f/g(\tilde k)$. This by itself is sufficient to
establish  a current proportional to $E^{d/2}$ in dimensions 
$d<3$.
For $d > 3$ the fixed point is Gaussian and the necessity of including
dangerously irrelevant scattering processes will modify this scaling.

We can make progress on the actual solution by making two 
simplifications valid for the leading order (in $1/N$) computation 
of the current: we can ignore the mass
renormalization and eliminate the
second equation in favour of a local source term\cite{Local}.
With these we find the greatly simplified and soluble equation
\begin{eqnarray}
{\bf E}.\frac{\partial f({\bf k})}{\partial {\bf k}}
&=&
-\gamma_{\bf k} f({\bf k})
+e^{-\pi \epsilon_{{\bf k}}^2/E} \delta ({\bf k.E}/E^2)
\label{Boltzman2}
\end{eqnarray}
with $\gamma_{\bf k}$ and $m(E)$ given by Eq.~(\ref{Boltzman}) in the
case of steady-state distributions.

The current carried in the steady-state is given by
\begin{eqnarray}
{\bf j} 
&=& 
2 \int \frac{d {\bf k}}{(2 \pi)^2}
\frac{k_{\parallel}}{\epsilon_{\bf k}} f({\bf k}) \ \ ,
\end{eqnarray}
with an additional term involving $g({\bf k})$ being subdominant in $1/N$.
Upon rescaling, this reduces to a form that, in 2-dimensions,
is proportional to the electric field. The constant of proportionality
is universal and may be calculated in the $1/N$ limit to be
$\sigma=(N \pi/8)e^{*2}/h$\cite{GappedPhase}, where $N=2$ in the 
physical system.

\paragraph*{Dissipation and higher orders}

The steady state that we have described involves a balance between
pair production, their acceleration by the field and relaxation
due to scattering. The latter processes need to relax the current,
the number of charge carriers, as well as the energy. The first
two require processes present in the Hamiltonian but energy
relaxation is present only at leading order where the
infinitely many orthogonal modes act as a heat sink. In order
that we be able to go to higher orders in $1/N$ we need to 
understand how that problem is to be dealt with. Qualitatively
we expect to mimic the logic of linear response theory, where
Joule heating is an $O(E^2)$ process which can be dealt with
without invalidating the $O(E)$ result one computes. In our
case the leading current response is $O(E^{d/2})$ but the
Joule heating ${\bf j \cdot E}$ is still down by a factor of
$E$ so the same logic is {\it prima facie} applicable. Explicitly,
this can be done by constructing a solution that carries a heat
current to the boundaries of the sample, e.g. transverse to
the direction of electric current flow\cite{Tremblay1979}. Roughly, this can be
thought of as a local ``effective temperature'' $T_{\rm eff}
(y)$ which drives a heat current $\kappa \nabla^2 T_{\rm eff} (y)$.
By scaling, $T_{\rm eff} \propto \sqrt{E}$. Requiring that
the variation in $T_{\rm eff}$ be less than its mean value and
factoring in the scaling form of $\kappa$ leads to the 
conclusion that such a solution can be constructed for a system
with transverse dimension, $L_y$, and electric fields satisfying
$L_y \sqrt{E}\ll 1$. We note that such a restriction is not unusual, a similar
construction for a ordinary metal at finite temperature also yields
a bound on system size and electric field\cite{Tremblay1979}. 
While the estimate that we give would apply to a generic critical point
with $z=1$, our particular
system is an even better bet for such a construction as $\kappa$
is infinite at all $T$ due to an absence of energy current relaxation. 

\paragraph*{Bose gases}

While this work was motivated by the fundamental question of
understanding the non-linear quantum-critical states, we would be
remiss if we did not note that the Mott transition described
by (1) has been observed, remarkably enough,
in atomic Bose-Einstein condensates placed in optical lattices of 
variable depth\cite{Greiner1998,Jaksch1998}.
In this system, the
role of the electric field is played by an intensity gradient in the
optical field, or alternatively by an acceleration of the optical
lattice. The universal current predicted within our analysis amounts
to a steady flow of matter proportional to the acceleration of the
optical lattice. Field gradients and accelerations in these systems
are easily made large in the sense of our theoretical discussion.
Whether the non-linear response discussed here can be observed
given the complications of the confining potential and the
differing requirements of equilibration in these systems appears
to be a fit topic for further theoretical and experimental work.

We would like to thank Leon Balents and Eduardo Fradkin for early discussions on the problem and Subir Sachdev for a critical reading of the manuscript. We  thank NSF grants DMR-9978074 and 0213706, the David and Lucile Packard Foundation and the Royal Society  for financial support.


\begin{thebibliography}{99}
\bibitem{Sachdev1999} S. Sachdev, {\it Quantum Phase Transitions}, CUP (1999), see also 
S. L. Sondhi {\it et al}, 
Rev. Mod. Phys. {\bf 69}, 315-333 (1997)

\bibitem{Damle1997}
K. Damle and S. Sachdev,
Phys. Rev. B56, 8714 (1997)

\bibitem{Kluger1991}
See for example Y. Kluger, J. M. Eisenberg, B. Svetitsky, F. Cooper and E. Mottola,
Phys. Rev. Lett.67, 2427
(1991) and references therein.

\bibitem{Cha1991}
M.-C. Cha, M. P. A. Fisher, S. M. Girvin, Mats Wallin and
A. P. Young, 
Phys. Rev. B44, 6883 (1991).

\bibitem{Dalidovich2004}
Our problem is rather different from the interesting work by
Dalidovich and Phillips in 
Phys. Rev. Lett. 93, 027004
(2004). In that work a dissipative bath is part of the fixed point
action and so the existence of nonlinear response is immediate, much
as in stochastic models of classical critical dynamics, e.g.
A. T. Dorsey, 
Phys. Rev. B 43, 7575 (1991). Hamiltonian actions, such
as the one we consider, present trickier questions.

\bibitem{Distributions}
Strictly, the distribution functions are defined through a Wigner transformation of the regular and anomalous Green's functions and have the forms
$
f({\bf k},t)=\int d^2q \langle a^{\dagger}({\bf k}+{\bf q}/2,t)a({\bf k}-{\bf q}/2,t) \rangle /(2 \pi)^2
$
and
$g({\bf k},t)=\int d^2q \langle b^{\dagger}(-{\bf k}-{\bf q}/2,t)a({\bf k}-{\bf q}/2,t) \rangle /(2 \pi)^2
$.
We assume this form implicitly in the main text, whilst retaining the simpler notations
$
f({\bf k},t)=\langle a^{\dagger}({\bf k},t)a({\bf k},t) \rangle$
and
$g({\bf k},t)=\langle b(-{\bf k},t)a({\bf k},t) \rangle$.

\bibitem{Zener1932}
C. Zener, Proc. Roy. Soc A137, 696 (1932), L. Landau, Sov. Phys. 1,
89 (1932); L. D. Landau and E. M. Lifshitz, Quantum Mechanics,
Pergamon, Oxford (1965).

\bibitem{Sachdev1998}
S. Sachdev, 
Phys. Rev B57, 7157 (1998).

\bibitem{gapON}
The dominant scattering at this order for E-field coupled modes is
from other E-field coupled modes; at zero-temperature and to lowest
order in $1/N$ there 
are no fluctuations in the other $N-1$ copies. Because of this, the
dynamically generated gap $m(E)$ is $O(1/N)$ rather than $O(1)$ as
would be the case if it were generated by thermal equilibrium
scattering.

\bibitem{Local}
The terms $\gamma {\cal R}e g $ in the first equation and $\gamma f$ in
the second are also neglected. This is justified by considering the order of magnitude of each term in the Boltzmann Eqs.(6) at differing values of $k_{\parallel}$. The neglected terms are subdominant in $1/N$ throughout the range of $k_{\parallel}$.

\bibitem{GappedPhase}
Interestingly, if one calculates the response in the gapped phase
(adding a constant term, $\Delta^2$ to the renormalized gap so that
$m^2(E=0)=\Delta^2$), one also finds an apparent universal conductivity. This
was first recognised in the linear response at thermal equilibrium by
Denis Dalidovich and Philip Phillips in 
Phys. Rev. B{\bf 64}, 052507 (2001). It
occurs due to a cancelation in contribution 
to the conductivity of the
exponentially small density of carriers with the correspondingly
small scattering between them. 

\bibitem{Tremblay1979}
A.-M. Tremblay, B. Patton, P. C. Martin and P. F. Maldague,
Phys.Rev. A19, 1721 (1979).

\bibitem{Greiner1998}
M. Greiner, O. Mandel, T. Esslinger, T. W.  Hansch, I. Bloch,
Nature 415, 39 (2002).

\bibitem{Jaksch1998} 
D. Jaksch, et al.,
Phys. Rev. Lett. 81, 3108 (1998)  



\end{thebibliography}
\end{document}